\documentclass{aastex631}

\usepackage{physics}
\usepackage{graphicx}
\usepackage[T1]{fontenc}
\usepackage{amsmath}
\usepackage[utf8]{inputenc}
\renewcommand{\baselinestretch}{1.3}
\usepackage{xcolor}

\newcommand{\ua}{\textbf{u}_{\text{A}}}
\newcommand{\da}{\boldsymbol\nabla\cdot\textbf{u}_{\text{A}}}
\newcommand{\dap}{\boldsymbol\nabla'\cdot\textbf{u}_{\text{A}}'}
\newcommand{\uh}{\textbf{u}}
\newcommand{\divv}{\boldsymbol\nabla\cdot\textbf{u}}
\newcommand{\dvp}{\boldsymbol\nabla'\cdot\textbf{u}'}
\newcommand{\bnabla}{\boldsymbol{\nabla}}
\newcommand{\bu}{\mathbf{u}}
\newcommand{\bua}{\mathbf{u_A}}

\usepackage{makecell} 
\usepackage{booktabs} 

\usepackage[linecolor=red!30]{todonotes}
\tikzstyle{notestyleraw} = [
draw=black!0,
line width=0.8pt,
fill=red!10,
text=black,
text width=3cm - 1.6 ex - 1pt,
inner sep=0.8 ex
]

\begin{document}

\shorttitle{The Energy Cascade Rate in Supersonic MHD Turbulence}
\title{The Energy Cascade Rate in Supersonic Magnetohydrodynamic Turbulence}

\correspondingauthor{Gonzalo Javier Alvarez}
\email{galvarez@df.uba.ar}
\author[0009-0006-2711-2602]{Gonzalo Javier Alvarez}
\affiliation{Universidad de Buenos Aires, Facultad de Ciencias Exactas y Naturales, Departamento de Física, Ciudad Universitaria, 1428 Buenos Aires, Argentina}
\affiliation{CONICET - Universidad de Buenos Aires, Instituto de Física Interdisciplinaria y Aplicada (INFINA), Ciudad Universitaria, 1428 Buenos Aires, Argentina}

\author[0000-0002-0786-7307]{Pablo Dmitruk}
\affiliation{Universidad de Buenos Aires, Facultad de Ciencias Exactas y Naturales, Departamento de Física, Ciudad Universitaria, 1428 Buenos Aires, Argentina}
\affiliation{CONICET - Universidad de Buenos Aires, Instituto de Física Interdisciplinaria y Aplicada (INFINA), Ciudad Universitaria, 1428 Buenos Aires, Argentina}

\author[0000-0001-5372-6882]{Branislav Rabatin}
\affiliation{Department of Physics, Florida State University, Tallahassee, Florida, USA}

\author[0000-0001-6661-2243]{David C. Collins}
\affiliation{Department of Physics, Florida State University, Tallahassee, Florida, USA}

\author[0000-0002-1272-2778]{Nahuel Andr\'es}
\affiliation{Universidad de Buenos Aires, Facultad de Ciencias Exactas y Naturales, Departamento de Física, Ciudad Universitaria, 1428 Buenos Aires, Argentina}
\affiliation{CONICET - Universidad de Buenos Aires, Instituto de Física Interdisciplinaria y Aplicada (INFINA), Ciudad Universitaria, 1428 Buenos Aires, Argentina}

\begin{abstract}
Three-dimensional direct numerical simulations (DNS) are implemented to investigate the energy cascade rate in compressible isothermal magnetohydrodynamic (MHD) turbulence. Utilizing an exact law derived from the Kármán-Howarth equation, we examine the contributions of flux and non-flux terms to the cascade rate across a broad range of sonic and Alfvénic Mach numbers, from subsonic to supersonic regimes and varying mean magnetic fields. Cascade rates are computed using on-grid 3-D decomposition and two plasma increment approaches: signed and absolute values.
Anisotropy induced by strong magnetic fields is analyzed through angular-dependent scaling of the cascade terms. Moreover, the increment calculation method significantly influences the relative contributions of flux and non-flux terms, with absolute methods tending to overestimate the latter. These findings extend current studies of compressible turbulence and offer critical insights into energy transfer mechanisms relevant to many astrophysical phenomena.

\end{abstract}

\keywords{Turbulence --- Magnetohydrodynamics --- Compressible --- Supersonic}

\section{Introduction}\label{sec:intro}
Turbulence is a highly complex and irregular state of motion that evolves both in space and time. Despite its chaotic behavior, turbulent flows exhibit organized structures across a range of sizes and lifetimes, which interact non-linearly to produce the so-called nonlinear energy cascade \citep[see, e.g.,][]{R1922}. A cornerstone of turbulence theory lies in the study of incompressible hydrodynamic (HD) turbulence, where \citet{vkh1938} derived an exact expression for the energy cascade rate, $\varepsilon$, under the assumptions of infinite Reynolds number, time stationarity, and homogeneous and isotropic turbulence. This result, formalized in the von Kármán-Howarth equation, remains a foundational aspect of incompressible turbulence theory. Among its many consequences is the derivation of the Kolmogorov spectrum in the inertial range for the kinetic energy \citep{K1941a,K1941b,MY1975}. Notably, the von Kármán-Howarth equation establishes a connection between large-scale quantities, such as velocity field increments, and small-scale dissipation rates, independent of the fluid's specific dissipation mechanisms. The von Kármán-Howarth equation has been extended to magnetized plasmas in different theoretical derivations \citep[see,][]{P1998a,P1998b}. Under the same assumptions as in incompressible HD turbulence, \citet{P1998a} derived an exact relation for fully developed incompressible magnetohydrodynamic (IMHD) turbulence. This relation, formulated in terms of the divergence of a flux-like term, involves two-point third-order structure functions and leads to the so-called 4/3 law for IMHD turbulence. Numerous numerical studies have validated this relation under a variety of system parameters and conditions \citep[e.g.,][]{SV2002, Mi2003}. Furthermore, this relation has been successfully applied using observational data, where it has been utilized to estimate the incompressible energy cascade rate in the solar wind \citep[e.g.,][]{SV2007,Mc2008,M2008,St2010,A2022}, demonstrating its relevance beyond theoretical and numerical contexts.

The solar wind is a supersonic plasma originating from the expansion of the solar corona into the heliosphere \citep[e.g.,][]{BC2013,P2019}, which provides a natural laboratory for studying MHD and plasma turbulence. While most theoretical advancements in MHD turbulence have relied on the incompressibility approximation, the inclusion of density fluctuations introduces significant modifications to the plasma dynamics \citep[see, e.g.,][]{Di2005, B2013, A2017b, B2021}. In particular, the solar wind can be characterized by its level of compressibility, anisotropy and its connection with the turbulent properties \citep{TM1995, Da2005, FRA2019}. In the context of compressible turbulence, \citet{B2013} derived an exact relation for isothermal and compressible magnetohydrodynamic (CMHD) turbulence using the compressible Els{a}sser variables. \citet{A2017b} derived an exact relation expressed in terms of plasma velocity, compressible Alfvén velocity, and plasma density. Their results revealed that the compressible energy cascade rate, $\varepsilon_C$, includes not only traditional flux-like terms but also novel non-flux contributions, which are presumably negligible within the inertial range \citep{H2017a,A2019b}. These theoretical results have been validated using numerical simulations and in situ observations in subsonic, quasi-incompressible and strongly compressible regimes \citep{B2016c,H2017a,A2019b,B2020,A2020,A2021,A2023,R2022,B2023,R2024}. Using three-dimensional (3-D) direct numerical simulations (DNSs), \citet{A2018} investigated the energy cascade rate and the relative importance of each term in isothermal compressible MHD turbulence, for different subsonic Mach numbers $M_S$ and different magnetic guide fields strengths $B_0$. The authors found that the dominant contribution to the cascade comes from the flux-like term, which depends weakly on the magnetic guide field, unlike the non-flux terms whose absolute values increase significantly with $M_S$ and $B_0$. \citet{B2023} estimated the incompressible and compressible MHD cascade rate at different heliocentric distances. Using in situ magnetic field and plasma observations provided by the Parker Solar Probe (PSP) mission, the authors compared two exact relations for compressible isothermal and polytropic MHD turbulence, respectively. Their results show an increase of the incompressible and compressible energy cascade rates as they approach to the Sun, with essentially the same isothermal and polytropic cascade rates in the range of compressibility and anisotropy observed. Finally, the compressible exact relation derived by \citet{A2017a} were subsequently validated and extended by \citet{S2021}, who generalized the analysis to include compressible and polytropic plasmas. This extension broadened the applicability of the theory and further highlighted the complex dynamics underlying compressible turbulence. 

In situ observations of the solar wind reveal that proton temperatures decrease with radial distance from the Sun more slowly than predicted by adiabatic expansion models \citep[e.g.,][]{M2011}, highlighting the longstanding "solar wind heating problem". This challenge remains a central topic in space physics \citep{Sm2024}. One potential explanation lies in the turbulent energy cascade, which transfers energy to dissipation scales where electromagnetic energy is converted into plasma heating and/or particle acceleration \citep{BC2013}. In this framework, large-scale MHD turbulence serves as an energy reservoir, channeling energy to smaller scales where dissipation occurs through various kinetic processes \citep[see,][]{V2016,Sc2022,P2023}. Despite this theoretical understanding, the role of compressibility—particularly in supersonic plasmas dominated by significant density fluctuations—remains poorly understood. Furthermore, compressible regimes, such as those found in planetary magnetospheres, the interstellar medium, and star-forming regions, exhibit prominent density fluctuations that strongly influence the nonlinear dynamics \citep[e.g.,][]{W1970,E2004}. In this sense, \citet{F2020} investigated the role of supersonic turbulence using two-point correlations functions in an exact relation that contains a single flux-like term (reminiscent of the incompressible one) and a source term. The authors found that the flux term contributes positively to the total energy cascade rate, which was interpreted as a direct cascade amplified by the compression. Also, their numerical results reveal the existence of different turbulent regimes separated by the sonic scale, which determines the scale over which the non-negligible source modifies the scaling of the flux.
In another context, \citet{Gz2016} has shown the important influence of compressibility in the acceleration of charged particles through numerical simulations of test particles under CMHD versus IMHD fields.


In this paper, we investigate the energy cascade rate in isothermal CMHD turbulence using three-dimensional 3-D DNSs. Through a parametric study varying the sonic Mach number and the Alfvén Mach number, we explore a wide range of compressibilities, spanning from subsonic to supersonic regimes, as well as different levels of anisotropy. The paper is organized as follows: Section \ref{sec:theory} introduces the compressible MHD equations, the exact relation for fully developed turbulence, and its different components. Section \ref{sec:code} describes the numerical code and the simulation setup. Section \ref{sec:results} presents the main numerical results. Finally, Section \ref{sec:conclusions} summarizes the key findings.

\section{THEORY}
\label{sec:theory}

\subsection{Compressible MHD model}
\label{sec:model}

The governing equations for compressible magnetohydrodynamics (CMHD) include the continuity equation for mass density, the momentum equation for the velocity field with the Lorentz force, the induction equation for the magnetic field, and Gauss’s law for magnetism \citep[see, e.g.,][]{F2014,A2017a}. Following \citet{M1987}, the system can be expressed in terms of the velocity field $\mathbf{u}$, the Alfvén velocity $\mathbf{u}_A = \mathbf{B}/\sqrt{4\pi \rho}$ (where $\mathbf{B}$ is the magnetic field and $\rho$ the mass density), the scalar pressure $P$, and the internal energy $e$. Assuming an isothermal equation of state, the pressure is given by $P = c_s^2 \rho$, where $c_s$ is the constant sound speed. Under this assumption, the internal compressible energy can be written as $e \equiv c_s^2 \ln(\rho/\rho_0)$, with $\rho_0$ a reference density:

\begin{align}\label{1} 
    & \frac{\partial e}{\partial t}  = - \uh\cdot\boldsymbol\nabla e-c_s^2\boldsymbol\nabla\cdot\uh , \\
    \label{2} 
    &\frac{\partial \textbf{u}}{\partial t} = -\uh\cdot\boldsymbol\nabla\uh  + \ua\cdot\boldsymbol\nabla\ua - \frac{1}{\rho}\boldsymbol\nabla(P+P_M) - \ua(\da) + \textbf{f}_k  + \textbf{d}_k , \\ 
    \label{3} 
    &\frac{\partial \ua}{\partial t} = - \uh\cdot\boldsymbol\nabla\ua + \ua\cdot\boldsymbol\nabla\uh -\frac{\ua}{2}(\divv) + \textbf{f}_m + \textbf{d}_m , \\  
    \label{4} 
    &\ua\cdot\boldsymbol\nabla\rho = -2\rho(\da).
\end{align}
The model also includes a mechanical forcing term $\mathbf{f}_k$ and the curl of an electromotive forcing term $\mathbf{f}_m$, along with small-scale dissipation terms $\mathbf{d}_k$ and $\mathbf{d}_m$ acting on the velocity and magnetic fields, respectively. In addition, we define the magnetic pressure as $P_M = \rho u_A^2 / 2$.

\subsection{Exact  relation for CMHD turbulence}
\label{sec:cascade}

Assuming a fully developed homogeneous plasma turbulence, a stationary state, and a balance between forcing and dissipation, \citet{A2017b} derived a complete expression for the exact relation in isothermal and compressible MHD turbulence. In its compact form, the relation can be expressed as
\begin{align}
\label{exact_law}
    -2 \varepsilon_C = \frac{1}{2} \bnabla_\ell \cdot \mathbf{F}_C + \text{S}_C + \text{S}_H + \text{M}_\beta,
\end{align}
where $\varepsilon_C$ is the total compressible energy cascade rate, $\mathbf{F}_C$ is the total compressible flux term, and $\text{S}_C$, $\text{S}_H$, and $\text{M}_\beta$ are the so-called source, hybrid, and $\beta$-dependent non-flux terms, respectively. The detailed derivation of the complete compressible energy cascade expression can be found in \citet{A2017a}. A remarkable feature of these exact relations is that no prior assumption of isotropy is required, and they are independent of the type of dissipation mechanism acting on the plasma (it is only assumed that dissipation acts on the smallest scales of the system) \citep[see also,][]{B2013,A2016b,A2016c}. 

To express the structure and two-point correlation functions, we introduce operations involving displaced field quantities. For clarity, we adopt a prime notation: $\mathbf{r}' = \mathbf{r} + \boldsymbol{\ell}$, where $\boldsymbol{\ell}$ is the displacement vector between two points. For instance, $\uh'$ represents the value of $\uh$ at the displaced position, $\uh(\mathbf{r}')$. Using this notation, we define the field increments as $\delta \uh \equiv \uh' - \uh$ and the local mean as $\bar{\uh} \equiv (\uh' + \uh)/2$, applicable to both scalar and vector fields. Additionally, the angular bracket $\langle \cdot \rangle$ indicates an ensemble average. In practice, rather than averaging over multiple realizations, we assume ergodicity \citep{Ba1953}, allowing us to compute results from a single realization by averaging over a sufficiently large spatial domain, such as all grid points within a DNS box.

\subsubsection{Flux terms}
The compressible flux terms, i.e., those terms that can be written as a function of structure functions in {Eq.~}\eqref{exact_law} are associated with the energy flux and can be decomposed as $\textbf{F}_\text{C}=\textbf{F}_{1C}+\textbf{F}_{2C}$:
\begin{align}\label{f1c}
	\textbf{F}_{1C} \equiv &~\big\langle[(\delta(\rho\uh)\cdot\delta\uh+\delta(\rho\ua)\cdot\delta\ua\big]\delta\uh - [\delta(\rho\uh)\cdot\delta\ua+\delta\uh\cdot\delta(\rho\ua)]\delta\ua\big\rangle,  \\ \label{f2c}
	\textbf{F}_{2C} \equiv & ~2\langle\delta e\delta\rho\delta\uh\rangle,
\end{align}
where $\textbf{F}_{1C}$ term is the inertial term, which is a compressible generalization of the third-order structure function derived by \cite{MY1975} for incompressible MHD, and the term $\textbf{F}_{2C}$ is a purely compressible contribution, without corresponding analogy in incompressible models. 

For incompressible MHD, the flux term can be recovered by neglecting density fluctuations (i.e., setting $\rho = \rho_0 = 1$). In this case, the flux term is given by\begin{align}\label{fi} \textbf{F}_I = \langle \left[ (\delta \bu)^2 + (\delta \bua)^2 \right] \delta \bu - 2 (\delta \bu \cdot \delta \bua ) \delta \bua \rangle. \end{align}From these definitions of the flux terms it is possible to compute the compressible energy cascade rate, $\varepsilon_{F_C}$, and the incompressible cascade, $\varepsilon_{F_I}$ for each direction of the increments.

\subsubsection{Non-flux terms}
The non-flux terms can be categorized into distinct components based on their mathematical properties and physical roles. The source terms, $S_C$, involve two-point correlation functions and are directly proportional to the divergence of the velocity fields, acting as a source or sink for the mean energy cascade rate within the inertial range. The hybrid terms, $S_H$, can be expressed either as a flux term or a source term; however, unlike $F_C$, they cannot be rewritten purely in terms of field increments when formulated as a flux term. Lastly, the $\beta-$dependent terms, $M_\beta$, originate from the magnetic pressure gradient and depend on the plasma $\beta$ parameter, which quantifies the ratio of plasma pressure to magnetic pressure \citep[see,][]{A2017b}. Each of the non-flux terms may be written as follows:

\begin{align}\nonumber
	  \text{S}_\text{C} \equiv&~ \langle[R_E'-\frac{1}{2}(R_B'+R_B)](\divv)+[R_E-\frac{1}{2}(R_B+R_B')](\dvp)\rangle \\ \label{term_source}
		&+\langle[(R_H-R_H')-\bar{\rho}(\uh'\cdot\ua)](\da)+[(R_H'-R_H)-\bar{\rho}(\uh\cdot\ua')](\dap)\rangle, \\ \nonumber
	  \text{S}_\text{H} \equiv&~ \langle\big(\frac{P_M'-P'}{2}-E'\big)(\divv)+\big(\frac{P_M-P}{2}-E\big)(\dvp)\rangle  + \langle H'(\da)+H(\dap)\rangle \\ \label{term_hybrid} &+\frac{1}{2}\langle\big(e'+\frac{u_\text{A}}{2}^{'2}\big)\big[\boldsymbol\nabla\cdot(\rho\uh)\big]+\big(e+\frac{u_\text{A}}{2}^2\big)\big[\boldsymbol\nabla'\cdot(\rho'\uh')\big]\rangle, \\ \label{term_beta}
	\text{M}_\beta \equiv&~ -\frac{1}{2}\langle\beta^{-1'}\boldsymbol\nabla'\cdot(e'\rho\uh) + \beta^{-1}\boldsymbol\nabla\cdot(e\rho'\uh') \rangle,
	\end{align}
where we have defined the total energy and the density-weighted cross-helicity per unit volume, respectively, as
\begin{align}\label{energy}
	E(\textbf{r}) \equiv &~\frac{\rho}{2}(\uh\cdot\uh+\ua\cdot\ua) + \rho e , \\
	H(\textbf{r}) \equiv &~\rho(\uh\cdot\ua),
\end{align}
and its corresponding two-point correlation function as
\begin{align}
	R_E(\textbf{r},\textbf{r}') \equiv&~ \frac{\rho}{2}(\uh\cdot\uh'+\ua\cdot\ua') + \rho e'  ,\\
	R_H(\textbf{r},\textbf{r}') \equiv&~ \frac{\rho}{2}(\uh\cdot\ua'+\ua\cdot\uh'),\\
    R_B(\textbf{r},\textbf{r}') \equiv&~\rho(\ua\cdot\ua')/2.
\end{align}Primed correlation functions implies an exchange in $\textbf{r}$ and $\textbf{r}'$.

Observationally, measuring the non-flux terms presents a significant challenge, as these terms require divergence calculations of the fields, which can only be performed using multi-spacecraft techniques. Even under optimal conditions, these techniques are subject to considerable uncertainties. Previous numerical investigations suggest that these terms are negligible compared to the flux terms in the inertial range, and would only be relevant in very specific regions of the plasma \citep{K2013, A2019a, F2020}. 

\subsection{The energy cascade rate components}
\label{sec:isotropic}

By following a similar procedure as in \cite{K2013}, assuming isotropy in the inertial range, a constant transfer energy $\varepsilon$  within this range, and integrating over a sphere of radius $\ell$ we get from Eq.~\eqref{exact_law}, the integrated contributions for the energy cascade along the box:
\begin{align}
      - \frac{4}{3} \varepsilon_C \ell = \text{F}_C + \text{Q}_S + \text{Q}_H + \text{Q}_M
\end{align}
where $\text{F}_C = \mathbf{F} \cdot  \boldsymbol{\hat\ell}$ is the parallel component in the increment direction and we define the integrated non-flux terms as:
\begin{align}
    \text{Q}_j(\ell) = \frac{1}{\ell^2} \int_0^\ell \text{S}_j(\ell^{'}) {\ell^{'}}^2 \dd \ell^{'} ,
\end{align}
where $\text{S}_j$ corresponds to $\text{S}_C$, $\text{S}_H$, and $\text{M}_\beta$, respectively, along with their associated non-flux integrated terms. Finally, the total energy cascade rate can be expressed as the contribution from the flux term, $\varepsilon_{F_C}$, and the non-flux terms, $\varepsilon_{Q_S}$, $\varepsilon_{Q_H}$, and $\varepsilon_{Q_M}$. The total non-flux cascade, sum of these terms, is denoted by $\varepsilon_{Q}$.

\section{Numerical Setup}
\label{sec:code}

\subsection{Numerical code}

\citet{WANG2009} extended a three-dimensional adaptive mesh refinement (AMR) MHD code based on the open-source cosmological AMR HD code ENZO \citep{BRYAN2014} to include magnetic fields. The solver uses the Dedner formulation \citep{DEDNER2002}, a conservative approach to the MHD equations that employs hyperbolic divergence cleaning to manage divergence constraints. This is achieved by coupling these constraints with conservation laws through Lagrange multipliers. The implementation of the method, originated in \citet[][]{MIGNONE20071427}, relies on the HLLD (Harten-Lax-van Leer Discontinuities) Riemann solver \citep{H1983}, combined with a piecewise linear reconstruction scheme for improved accuracy.

{\color{blue} In the numerical simulations, first presented in \citet{stalpes2024}, turbulence is reached and sustained by a stochastic mechanical forcing designed to drive large-scale fluid motions. This forcing is initialized in Fourier space, where the acceleration field $\hat{\mathbf{f}}(\mathbf{k},t)$ evolves in real space according to a generalized Ornstein-Uhlenbeck process \citep[see also,][]{F2010}. Energy is injected at a characteristic wavenumber corresponding to scales approximately half the domain size, ensuring the creation of large-scale eddies. Specifically, the forcing combines both compressive and solenoidal components, with the input power distributed such that the ratio of compressive to solenoidal amplitude is 2/3. To simulate isothermal conditions, we set the adiabatic exponent to $\gamma = 1.001$, ensuring the closure relation $P \simeq {c_S}^2 \rho$. This numerical setup allows the dissipation of kinetic energy to be redistributed into internal energy with minimal temperature variation, effectively approximating isothermal turbulence. The simulation boxes use periodic boundary conditions and a grid of $512^3$ points (or zones). The forcing method ensures statistically isotropic turbulence. At large scales, the compressive component dominates, driving shock formation and small-scale instabilities. In contrast, the solenoidal component generates vorticity and sustains large-scale eddies. The driving field is helicity-free to machine precision, ensuring no parity violation. Fully developed turbulence is reached after approximately five dynamical times, defined as $t_{dyn} = L_0 / M_S$, where $L_0$ is the forcing scale (half the box size). This stage is identified by the saturation of the mean energy. More specifically, all analyses presented in this work are performed at $t=10t_{dyn}$ to ensure a clear stationary state.}

{\color{blue} All simulations exhibit field spectra with a Kolmogorov like $k^{-5/3}$ scaling and a well resolved inertial range, as shown in \citet{stalpes2024}. Notably, the slopes of the velocity and magnetic field spectra do not depend of $M_S$, whereas the density spectrum slope shows a strong positive correlation. The integral scale is roughly half the box, as expected, and the Taylor scale is approximately 20 computational zones. It is worth mentioning that in compressible turbulence the definition of the Taylor microscale remains valid under compressive driving because shear dissipation remains the dominant mechanism for which the Taylor scale was originally formulated. Although compressive effects introduce bulk viscous dissipation, the ratio of shear to compressive dissipation,
\begin{align}
\frac{\langle(\nabla\times \mathbf{v})^2\rangle}{\langle(\nabla\cdot \mathbf{v})^2\rangle} \sim 10^4 ,
\end{align}
is very large in our simulations. This indicates that shear effects overwhelmingly dominate energy dissipation, which justifies the use of the Taylor microscale as the relevant dissipation length scale.}

Each simulation is characterized by two key parameters: the sonic and Alfvénic Mach numbers, defined as:
\begin{align}
    M_S &= \frac{u^{\text{rms}}}{c_S} \\
    M_A &= \frac{u^{\text{rms}}} {{B_0/\sqrt{\rho_0}}}
\end{align}
where the superscript "rms" denotes to root mean square value, $B_0$ is the mean magnetic field, $\rho_0$ is the mean mass density. The parameter space is explored by varying these parameters within the ranges of the sonic Mach numbers $$M_S = 0.5,1,2,3,4,5,6$$ and the Alfvénic Mach numbers $$M_A=0.5,1,2,$$ resulting in a total of $7\times3=21$ simulations. The mean mass density and sound speed are both set to unity, and these Mach numbers are controlled by the energy injection rate. Note that these parameters are target parameters and that actual values exhibit small fluctuations in each realization. 
\subsection{Method}
\label{sec:method}

To compute the correlation and structure functions, we calculated field increments in multiple directions using an SO(3) decomposition \citep[see, e.g.,][]{Ta2003}. This method ensures a homogeneous sampling of the simulation domain, thus avoiding the computational expense of 3-D interpolation outside the grid points \citep{Ma2010}. The decomposition is constructed using vectors (in grid-point units) such as (1,0,0), (1,1,0), (1,1,1), (2,1,0), (2,1,1), (2,2,1), (3,1,0), (3,1,1), combined with all possible permutations of indices and sign changes. This yields 73 unique directions that are used to calculate the structure functions. These functions are then averaged to derive the energy cascade terms as a function of the isotropic scale ($\ell$). It is important to note that not all directions have the same unit length, so in order to average all curves, we performed an interpolation to operate over the same spatial points. On the other hand, increments are carefully chosen to cover displacements up to half the box size, avoiding repeating terms due to the periodic boundary conditions.

A common methodology to improve statistical estimation of the energy cascade involves using the absolute value of the field increments. In this study, we calculate both signed and absolute field increments, enabling a direct comparison between these two methodologies for cascade estimation.

First, in Section \ref{sec:results:parameter_space}, we present the parameter space and analyze field fluctuations as functions of the simulation inputs. In Section \ref{sec:results:scales}, we examine the signed energy cascade terms as a function of scale. Section \ref{sec:results:flux} focuses on flux term analysis using the absolute increment method, which is widely employed in observational studies. Subsequently, we investigate the effects of compressibility and anisotropy in Sections \ref{sec:results:compressibility} and \ref{sec:results:anisotropy}, respectively. {\color{blue}In Section \ref{sec:results:flux_vs_nonflux} we provide a comprehensive comparison of the signed and absolute increment methods for both flux and non-flux contributions. Finally, in Section \ref{sec:results:relation_vs_injection}, we compare the estimated cascade to the energy injection rate.}


\section{Numerical Results}
\label{sec:results}

\subsection{Parameter space study}
\label{sec:results:parameter_space}

Table \ref{tab:tabla1} shows summary information about our 21 DNSs. It is worth mentioning that the sonic Mach number $M_S$ serves as the target value for root-mean-square velocity, ${u}^{\text{rms}}$, though statistical deviations are observed in each run. Similarly, Alfvénic to sonic Mach ratio $M_A/M_S$ sets the target value for the root-mean-square Alfvén velocity, ${u_A}^{\text{rms}}$. However, these statistical deviations become more pronounced at higher Mach numbers.

\begin{table}
\centering
\renewcommand{\arraystretch}{0.66} 
\begin{tabular}{|c|c c|c c c|}
 \hline
 \textbf{Run} & \multicolumn{2}{c|}{\textbf{Inputs}} & \multicolumn{3}{c|}{\textbf{Outputs}} \\
              & $M_S$ & $M_A$ & $B_0$ & $u^\text{rms}$ & ${u_A}^\text{rms}$ \\
\hline
 1  & $1/2$ & $1/2$ & $1.00$ & $0.57$ & $1.02$ \\ 
 2  & $1/2$ & $1$           & $0.50$ & $0.43$ & $0.57$ \\
 3  & $1/2$ & $2$           & $0.25$ & $0.36$ & $0.44$ \\[1ex]
 4  & $1$           & $1/2$ & $2.00$ & $1.01$ & $2.11$ \\ 
 5  & $1$           & $1$           & $1.00$ & $0.85$ & $1.20$ \\ 
 6  & $1$           & $2$           & $0.50$ & $0.71$ & $0.91$ \\[1ex]
 7  & $2$           & $1/2$ & $4.01$ & $2.01$ & $4.86$ \\
 8  & $2$           & $1$           & $2.00$ & $1.62$ & $2.75$ \\ 
 9  & $2$           & $2$           & $1.00$ & $1.42$ & $2.16$ \\[1ex]
 10 & $3$           & $1/2$ & $6.01$ & $2.51$ & $9.12$ \\
 11 & $3$           & $1$           & $2.99$ & $2.28$ & $5.05$ \\
 12 & $3$           & $2$           & $1.50$ & $1.95$ & $3.77$ \\[1ex]
 13 & $4$           & $1/2$ & $8.00$ & $4.68$ & $18.88$ \\
 14 & $4$           & $1$           & $4.00$ & $4.54$ & $10.51$ \\
 15 & $4$           & $2$           & $2.00$ & $3.99$ & $7.72$ \\[1ex]
 16 & $5$           & $1/2$ & $9.99$ & $5.86$ & $24.08$ \\
 17 & $5$           & $1$           & $5.00$ & $5.24$ & $14.86$ \\
 18 & $5$           & $2$           & $2.50$ & $4.97$ & $11.36$ \\[1ex]
 19 & $6$           & $1/2$ & $12.00$& $7.06$ & $30.32$ \\
 20 & $6$           & $1$           & $6.00$ & $6.92$ & $19.36$ \\
 21 & $6$           & $2$           & $3.00$ & $5.81$ & $13.64$ \\
\hline
\end{tabular}
\caption{DNSs parameters: sonic Mach numbers ($M_S$) and Alfvénic Mach numbers ($M_A$) are specified as inputs to the simulations. The mean magnetic field, $B_0$, along with the rms values of the velocity and Alfvén velocity, are computed within the simulation domain.}
\label{tab:tabla1}
\end{table}

\begin{figure}
\centering
\includegraphics[width=0.8\linewidth]{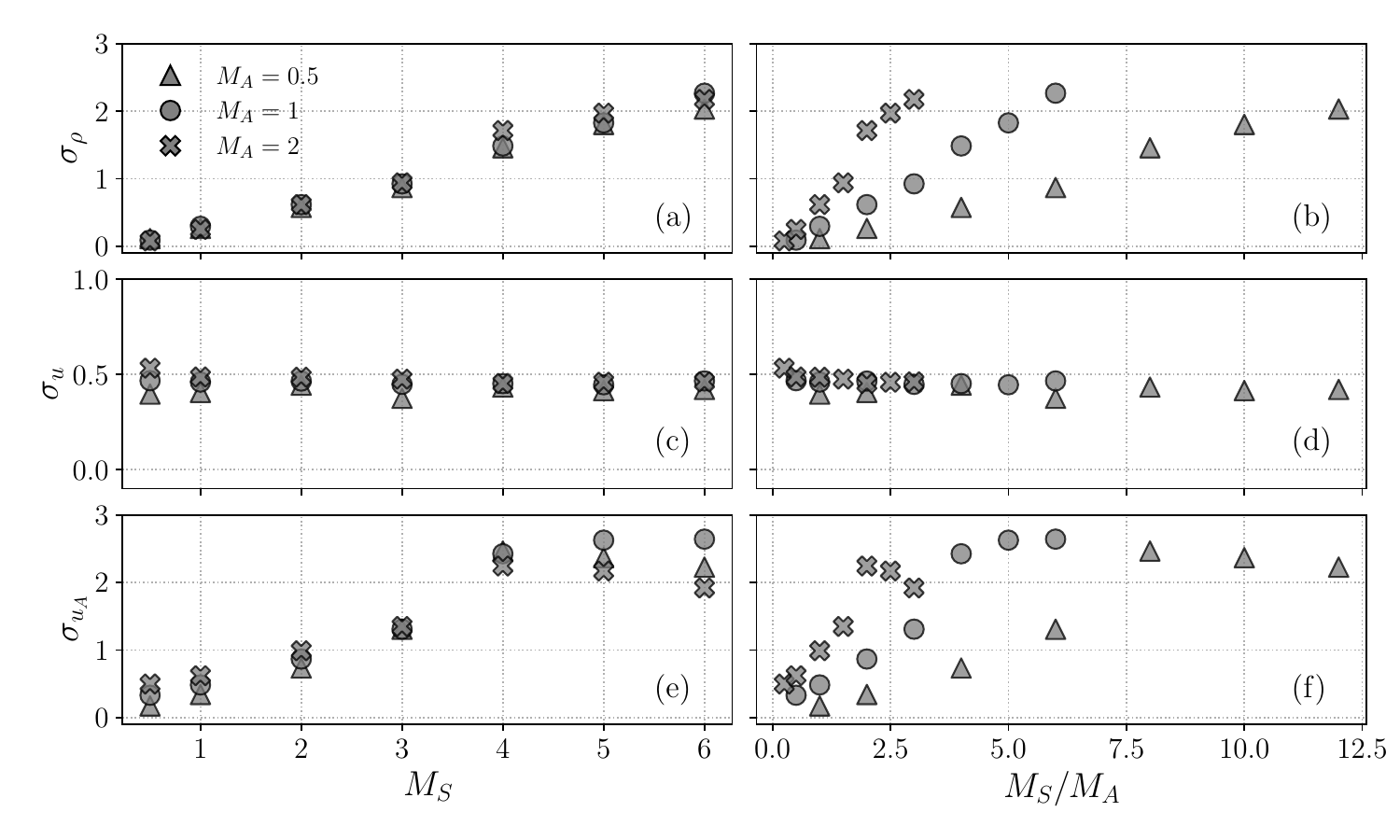}
\caption{Parameter space: standard deviation for density (a), velocity (b) and Alfvén velocity (c), respectively, as a function of the simulation inputs $M_S$ and the ratio $M_S/M_A$, which defines the mean magnetic field $B_0$ for each run.
\label{fig:fig1}}
\end{figure}

Figure \ref{fig:fig1} presents the standard deviation values for density, $\sigma_\rho,$ (a-b), velocity, $\sigma_u,$ (c-d), and Alfvén velocity, $\sigma_{u_A},$ (e-f) across the entire simulation domain as a function of the simulation inputs: the sonic Mach number ($M_S$) and the sonic to Alfvénic Mach number ratio ($M_S/M_A$), respectively. Different markers correspond to varying values of $M_A$. 

Our numerical results reveal that density fluctuations, $\sigma_\rho$, indicative of large-scale compressibility, and Alfvén velocity fluctuations depend directly on both $M_S$ and $M_A$, consistent with previous numerical works \citep[e.g.,][]{stalpes2024}. Specifically, plasma compressibility increases linearly with $M_S$ and shows only a weak dependence on $M_A$. In contrast, velocity fluctuations, $\sigma_{u}$, show no significant dependence on either $M_S$ or the ratio $M_S/M_A$, remaining nearly constant across all runs. Lastly, Alfvén velocity fluctuations, $\sigma_{u_A}$, show a linear trend for a fixed $M_A$ with a slope change around $M_S=3$ and $M_S=4$. In every case, these fluctuations increase until $M_S=3$, and then, they tend to decrease for larger compressibilities. These findings show the different scaling behavior of plasma density, velocity, and Alfvén velocity fluctuations in compressible MHD turbulence.
 
\subsection{Energy cascade rates vs scales}
\label{sec:results:scales}

\begin{figure}
\centering
\includegraphics[scale=0.6]{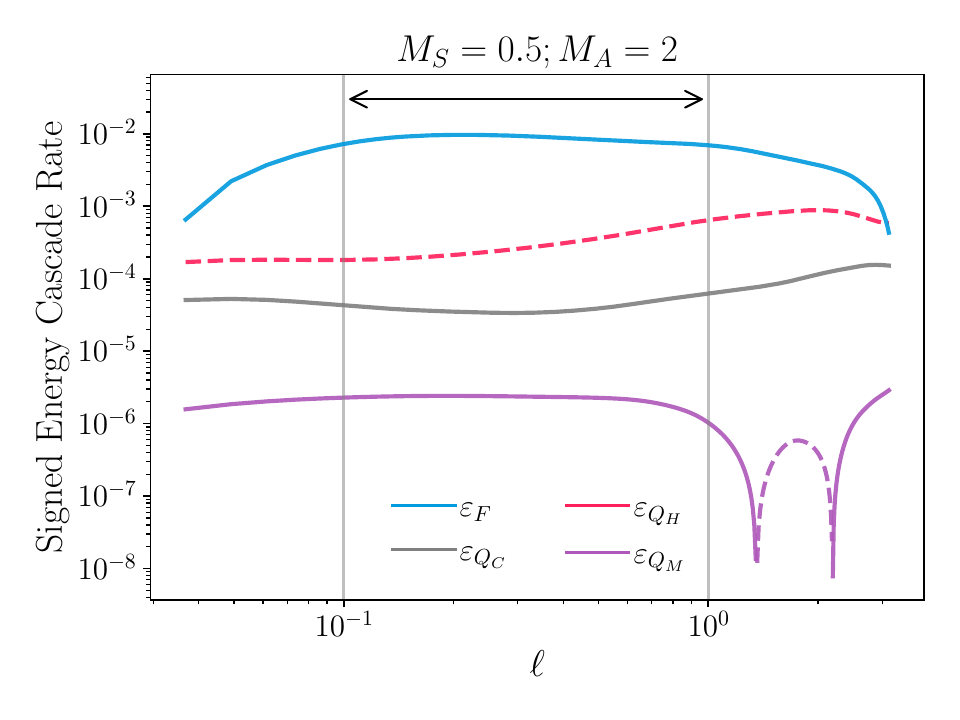}
\caption{Run 3: $M_S=0.5$ and $M_A=2$. Scaling of the energy cascade terms: flux ($\varepsilon_F$) and non-flux ($\varepsilon_{Q_C}$, $\varepsilon_{Q_H}$ and $\varepsilon_{Q_M}$).Dashed lines represent negative values. Vertical lines show the MHD scales (inertial range) used as averaging window.
\label{fig:fig2}}
\end{figure}

For all 21 simulations presented in Table \ref{tab:tabla1}, we computed the energy cascade terms using both the sign and absolute increments method, as mentioned. Figure \ref{fig:fig2} presents the total energy cascade rate and its components as a function of the scale for Run 3 as an example. The cascade due to the flux term, $\varepsilon_F$, dominates and defines the MHD inertial range, that is, the scales over which the total cascade rate remains approximately constant. In this case, the non-flux contributions to the total cascade (i.e., $\varepsilon_{Q_C}$, $\varepsilon_{Q_H}$, and $\varepsilon_{Q_M}$) are smaller and uniformly distributed across the MHD scales. To statistically investigate the behavior of the incompressible and compressible cascades across all simulations, we average the cascade rates and its different contributions within the MHD scale range, $\ell \in [10^{-1},10^{0}]$, to compute the mean cascade rate terms $\langle \varepsilon \rangle$ and, by using increments in absolute values, $\langle |\varepsilon| \rangle$.

\subsection{Statistical results: Flux terms}
\label{sec:results:flux}

In this Section, we investigate statistically both incompressible and compressible cascade contributions from the flux terms as a function of $M_S$ and $M_A$. More specifically, we study how compressibility (i.e., $\sigma_\rho$) and the level of anisotropy (i.e., $M_S/M_A\sim B_0$) and magnetic fluctuations in the plasma affect $\langle \varepsilon_{F_I} \rangle$ and $\langle \varepsilon_{F_C} \rangle$.

\subsubsection{Effect of compressibility}
\label{sec:results:compressibility}

\begin{figure}
\centering
\includegraphics[width=1\linewidth]{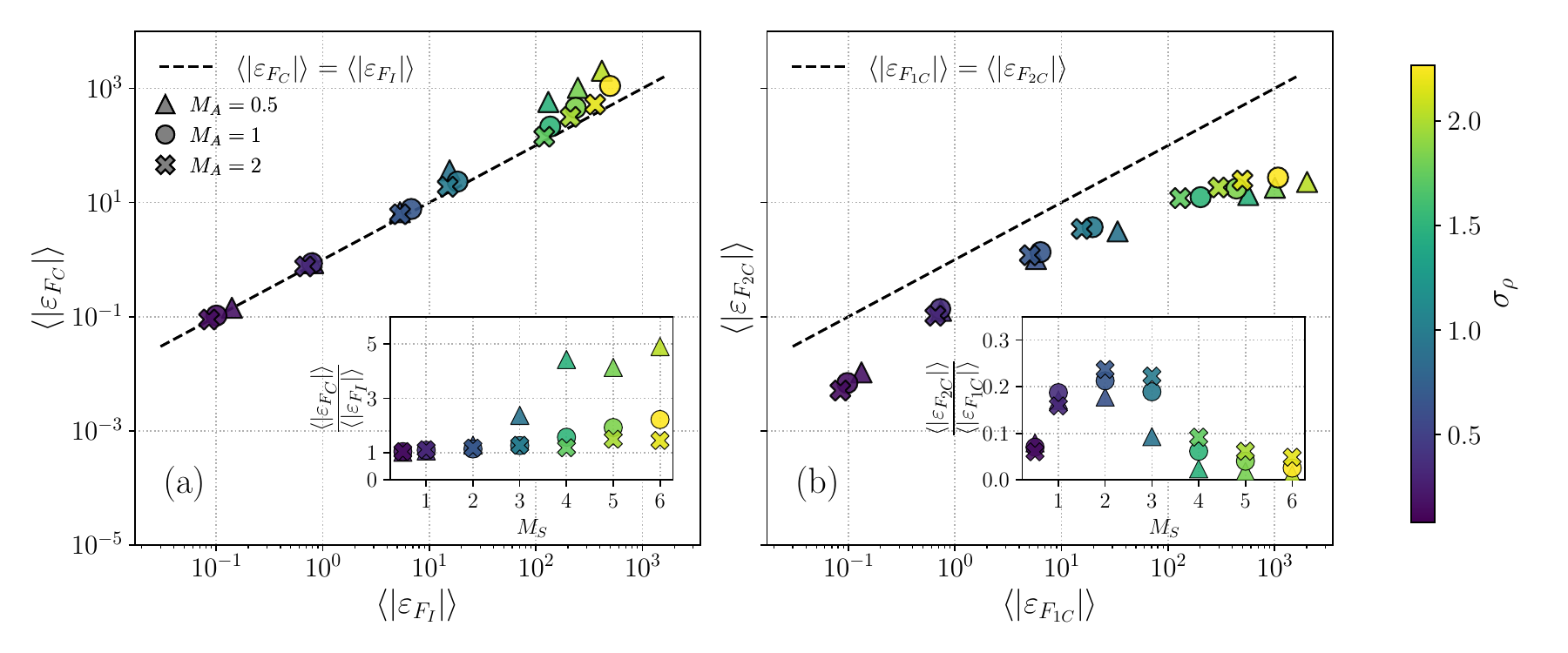}
\caption{(a) Compressible total flux $\expval{|\varepsilon_{F_C}|}$ as a function of the incompressible flux $\expval{|\varepsilon_{F_I}|}$. (b) Purely compressible term $\expval{|\varepsilon_{F_{2C}}|}$ as a function the term $\expval{|\varepsilon_{F_{1C}}|}$. Values shown are averaged at MHD scales and for absolute increments. Inset: ratio between the compressible and incompressible cascades as a function of the sonic Mach number $M_S$. The color bar corresponds to the compressibility level, $\sigma_\rho$.}
\label{fig:fig3}
\end{figure}

Figure \ref{fig:fig3}(a) compares the absolute values of the mean flux terms of the compressible and incompressible energy cascades as a function of compressibility (indicated by the color bar) which increases due to the sonic Mach number, $M_S$, as seen in Figure \ref{fig:fig1}(a). Different markers correspond to different Alfvénic Mach numbers ($M_A=0.5,1,2$). At low compressibility, regardless of $M_S$, $\langle |\varepsilon_{F_I}| \rangle$ and $\langle |\varepsilon_{F_C}| \rangle$ are identical, as expected from previous numerical and observational results \citep[see,][]{A2020,B2023}. However, as compressibility increases, the compressible cascade becomes larger than the incompressible cascade. Moreover, at a fixed $M_S$, decreasing $M_A$ (or equivalently, increasing the guide magnetic field $B_0$) leads to an enhancement of the compressible cascade. This is more evident in the inset of Figure \ref{fig:fig3}(a), which shows the ratio of compressible to incompressible cascades as a function of $M_S$.

In a similar format to Figure \ref{fig:fig3}(a), Figure \ref{fig:fig3}(b) compares the flux terms in the compressible cascade rate: the Yaglom-like term $\langle |F_{1C}| \rangle$ (see Eq.\ref{f1c}) and the purely compressible term $\langle |F_{2C}| \rangle$ (see Eq.\ref{f2c}). Two important features emerge. First, across different levels of compressibility, $\langle |F_{1C}| \rangle$ consistently dominates over $\langle |F_{2C}| \rangle$. Second, while $\langle |F_{1C}| \rangle$ significantly exceeds $\langle |F_{2C}| \rangle$ at both lower and higher plasma compressibilities, the two components approach similar values for intermediate values. The inset in Figure \ref{fig:fig3}(b) illustrates the ratio of the two compressible flux terms as a function of the sonic Mach number. There is no strong dependence on the guide magnetic field. However, for intermediate $M_S$ values, specifically between $M_S = 2$ and $M_S = 3$, we observe the highest values of $\langle |\varepsilon_{F_{2C}}| \rangle / \langle |\varepsilon_{F_{1C}}| \rangle$. At lower $M_S$ values, negligible contributions from $\langle |\varepsilon_{F_{2C}}| \rangle$ to the compressible cascade are expected. Interestingly, at higher sonic Mach numbers, our numerical results reveal a similar trend. This behavior is likely related to the existence of a supersonic regime, since as we increase the $M_S$, the rms velocity increases.

\begin{figure}
\centering
\includegraphics[width=\linewidth]{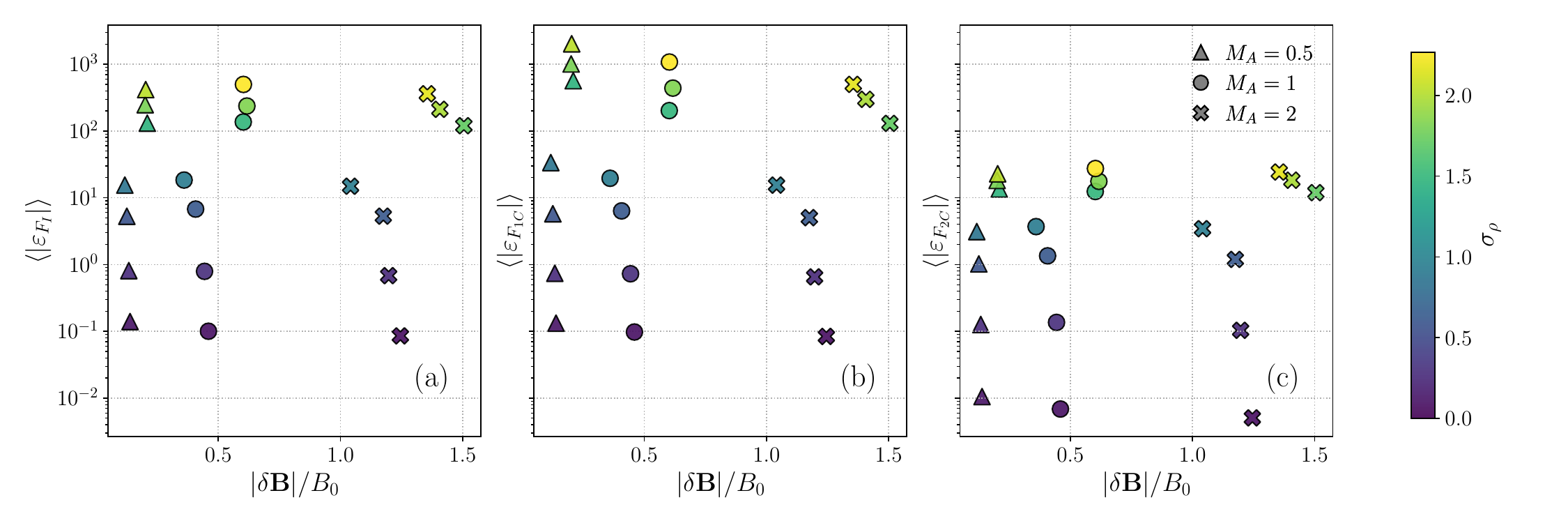}
\caption{Mean energy cascade rate from the incompressible cascade  terms $\expval{|\varepsilon_{F_{I}}|}$ (a), and the compressible terms: $\expval{|\varepsilon_{F_{1C}}|}$ (b) and $\expval{|\varepsilon_{F_{2C}}|}$ (c) as a function of magnetic field fluctuations. The color bar corresponds to the compressibility level, $\sigma_\rho$.}
\label{fig:fig4}
\end{figure}

Figure \ref{fig:fig4} (a-c) presents the compressible flux terms, $\langle |\varepsilon_{F_{1C}}| \rangle$ and $\langle |\varepsilon_{F_{2C}}| \rangle$, alongside the incompressible flux term, $\langle |\varepsilon_{F_{I}}| \rangle$, as functions of normalized magnetic field fluctuations, $\delta |\mathbf{B}| / B_0$. As the mean magnetic field increases, the system becomes more anisotropic, and the magnetic fluctuations are expected to decrease. Simulations with the highest mean magnetic field correspond to the regime of small magnetic fluctuations. In addition, the insets on Figures \ref{fig:fig3}(a) and \ref{fig:fig3}(b) demonstrate a notable change in behavior, which is associated with an increase in compressibility. From panels (a) and (b) we observe the previous results: The compressible cascade has a strong dependence on compressibility and on the level of anisotropy. However, it should be noted that the incompressible cascade expression in Eq. \eqref{fi} does not depend on the density fluctuation level, and the mean density is equal to one. Therefore, the dependence of this cascade in panel (c) is due only to the guide magnetic field $B_0$ and the increase in rms of the velocity field.

\subsubsection{Effect of anisotropy}
\label{sec:results:anisotropy}
To investigate the angular dependence of the compressible energy cascade rate, we transform the original Cartesian coordinate system into Spherical coordinates system. Assuming homogeneity around the $\hat{z}$ direction, defined as the direction of the mean magnetic field $\mathbf{B}_0$, we average over the azimuthal angle, $\varphi$, to compute a scale-dependent energy cascade for each polar angle, $\theta$ \citep[see,][]{Jiang2023}. This methodology enables the analysis of cascade behavior across polar angles, ranging from $0^\circ$ (i.e., parallel to $\mathbf{B}_0$) to $90^\circ$ (i.e., perpendicular to $\mathbf{B}_0$),
\begin{equation}
    \varepsilon(\ell; \theta_i) = 
    \frac{1}{2\pi} \int_{0}^{2\pi} \varepsilon(\ell, \theta_i, \varphi) \dd \varphi \approx \frac{1}{N_j} \sum_{j=1}^{N_j} {\varepsilon}(\ell; \theta_i, \varphi_j).
\end{equation}
In particular, for isotropic curves, we integrate over all the polar and azimuthal angles, applying the appropriate normalization factors:
\begin{equation}
    \varepsilon(\ell) = \frac{1}{4\pi} \int_{0}^{2\pi} \int_{0}^{\pi} \varepsilon(\ell; \theta,\varphi) \sin \theta \dd \theta \dd \varphi \approx \frac{\sum_{i=1}^{N_i} \sum_{j=1}^{N_j} \varepsilon(\ell; \theta_i, \varphi_j) \sin \theta_i}{N_j \sum_{i=1}^{N_i} \sin \theta_i }.
\end{equation}
Finally, this scale-dependent value is averaged over the MHD scales to obtain a single representative value for each run, $\expval{\varepsilon}_\text{MHD}$, or simply $\expval{\varepsilon}$, as presented in Section \ref{sec:results:flux}. To study the anisotropy in the flux terms, this analysis is performed using increments in absolute values to compute $\expval{|\varepsilon|}$ increasing statistics.

\begin{figure}
\centering
\includegraphics[width=.99\linewidth]{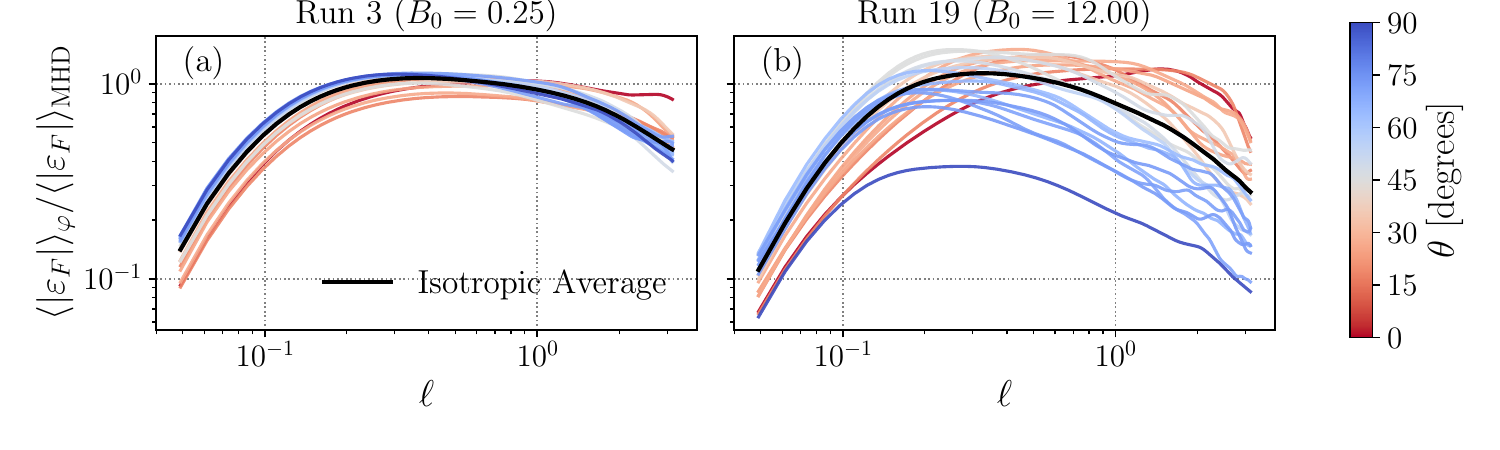}
\caption{Scaling of the absolute energy cascade rate from the flux-terms normalized by the mean value isotropic average in the MHD scales, $\expval{\varepsilon_F}_{\text{MHD}}$. Examples for: (a) Run 3, using $M_S = 0.5$ and $M_A =2$, and (b) Run 19, using $M_S=6$ and $M_A=0.5$. The color bar corresponds to the polar angle (where $\theta=0$ is the direction of the magnetic mean field). All curves are averaged across azimuthal angle $\varphi$. The black curve corresponds to the the averaged curve in both azimuthal and polar angle (isotropic average), $\expval{\varepsilon_F}_{\text{ISO}}$.
\label{fig:fig5}
}
\end{figure}

In Figure \ref{fig:fig5}, we compare Runs 3 and 9 to examine the impact of the guide magnetic field strength on the energy cascade across different polar angles $\theta$. Specifically, Figures \ref{fig:fig5}(a) and \ref{fig:fig5}(b) show the mean absolute value over the azimuthal angle of the energy cascade flux term, normalized to its value in the MHD scales, as a function of scale for Run 3 (a) and Run 9 (b). The color bar represents the polar angle. 

Our numerical results underscore the anisotropy introduced by the guide magnetic field and its role in modulating the energy cascade. Under weak mean field conditions, the cascade exhibits minimal variation with the polar angle, with all curves following a similar trend across different increment directions. In contrast, under a strong guide magnetic field, the cascade becomes more pronounced at larger scales (approximately $\ell \sim 1$) for increments parallel to the field direction. For increments perpendicular to the field, the cascade weakens and its peak shifts toward smaller scales (around $\ell \sim 0.2$). At intermediate polar angles (approximately $\theta = 30^\circ$ to $\theta = 60^\circ$), the cascade reaches its highest values, on scales around $\ell \sim 0.3$.

Finally, when averaging over all polar angles, the isotropic curve peaks within the MHD range, consistent with expectations for the inertial range. These results are consistent with previous numerical and observational studies of incompressible and anisotropic MHD turbulence \citep[e.g.,][]{A2022, Jiang2023}.

\subsubsection{Flux vs.~non-flux}
\label{sec:results:flux_vs_nonflux}

\begin{figure}
\centering
\includegraphics[width=.99\linewidth]{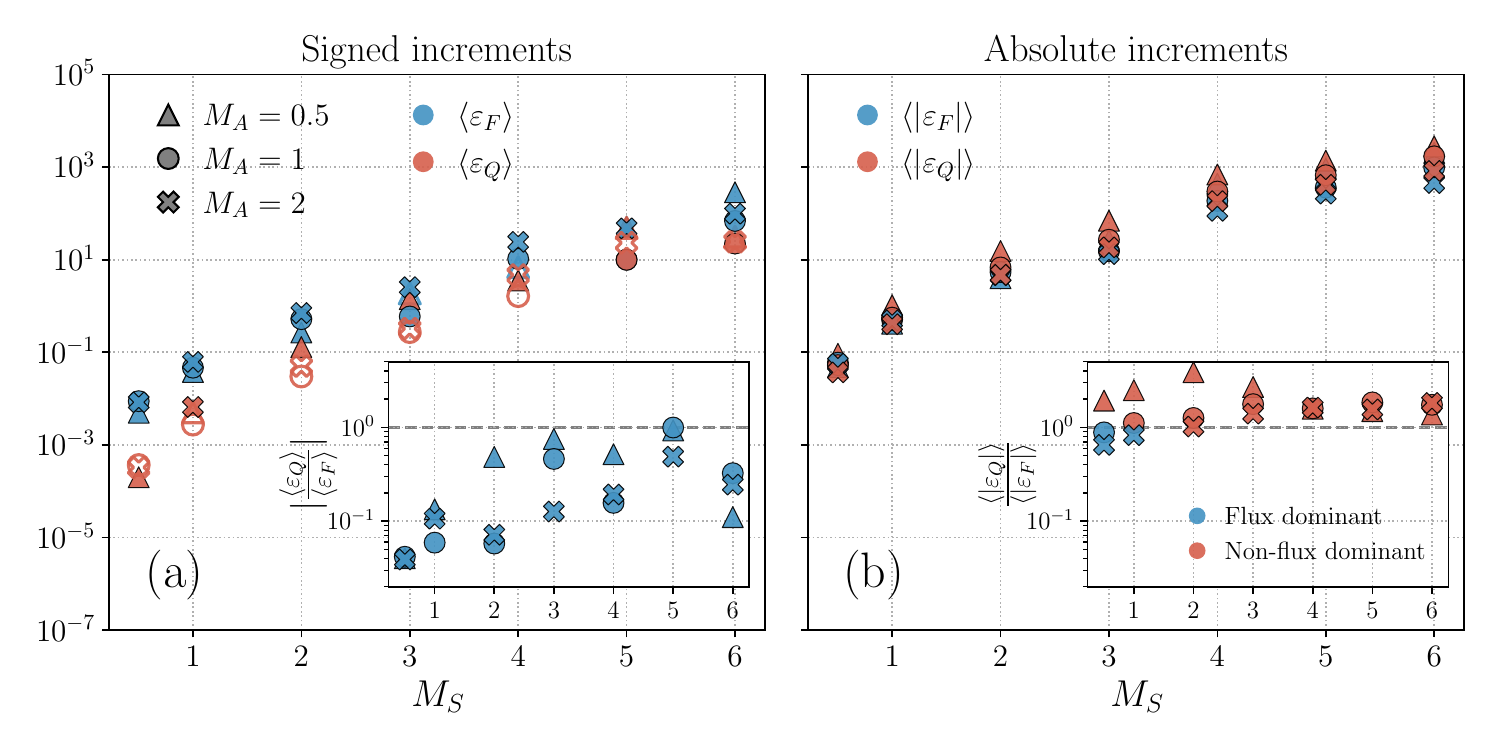}
\caption{(a) Mean of the signed compressible energy cascade rate at MHD scales from the flux terms, $\expval{\varepsilon_F}$ (blue), and the non-flux terms, $\expval{\varepsilon_Q}$ (red), across the 21 simulations. Unfilled markers represent negative values. Inset: absolute ratio of mean non-flux to flux contributions, with color indicating which is the dominant contribution (ratio lesser or greater than 1). (b) Same format as (a) using absolute increments instead of signed increments.
\label{fig:fig6}}
\end{figure}

Figure \ref{fig:fig6} shows the mean energy cascade flux (cyan) and non-flux (red) terms calculated using both signed (a) and absolute value increments (b) as a function of the sonic Mach number, $M_S$. The inset displays the absolute value of the flux-to-non-flux ratio to identify the dominant component for each run. Similarly to previous analyses, the markers represent different Alfvénic Mach numbers, $M_A$. As plasma compressibility increases, all cascade components grow. Notably, the significance of the non-flux contributions also increases for both signed and absolute value increments. However, when signed increments are used, the flux contributions consistently dominate within the MHD inertial range across all simulations. As $M_S$ increases, the relevance of the non-flux cascade grows; nevertheless, in most cases, the flux cascade remains dominant or at least comparable in magnitude to the non-flux contributions. In contrast, the use of absolute increments may be introducing spurious contributions, leading to a clear overestimation of the non-flux terms. This effect is particularly evident even in the subsonic and transonic regimes ($M_S = 0.5$ to $M_S = 1$), where a flux-driven cascade is expected. Remarkably, under these particular conditions, approximately half of the runs indicate non-flux dominance. An interesting trend emerges in this regime: as $M_A$ decreases (for a given $M_S$), the dominance of non-flux contributions increases, correlating with the presence of a stronger mean magnetic field. This observation highlights the importance of accurately distinguishing flux and non-flux contributions to understand energy transfer mechanisms in compressible turbulence, even for these small $M_S$ values.


\subsubsection{\color{blue}Exact Relation vs.~Energy Injection}
\label{sec:results:relation_vs_injection}

\begin{figure}
\centering
\includegraphics[width=\linewidth]{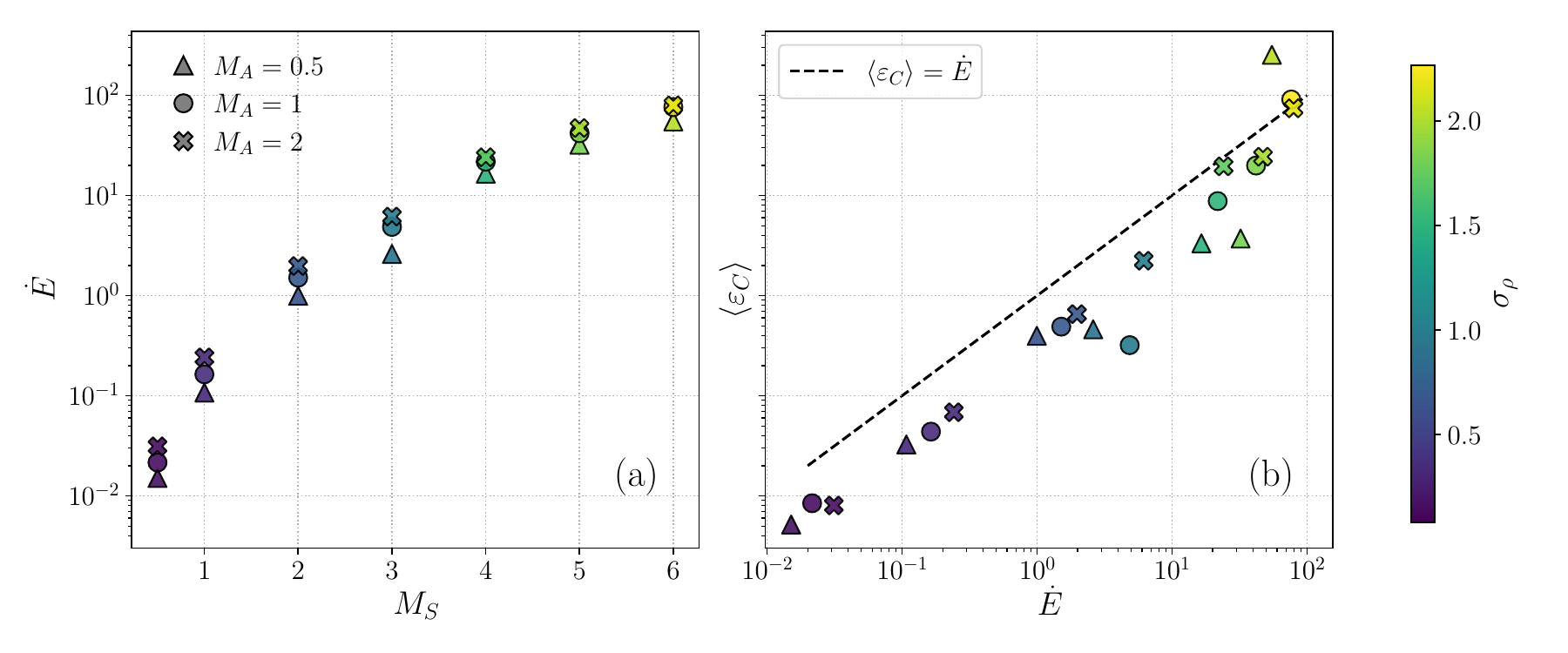}
\caption{\color{blue}{(a) Energy injection rate, $\dot E$, as a function of the mean compressible energy cascade rate, $\langle \varepsilon_C \rangle$. (b) Mean energy cascade rate as a function of the energy injection rate. The color bar corresponds to the compressibility level, $\sigma_\rho$.}}
\label{fig:fig7}
\end{figure}

{\color{blue} Figure \ref{fig:fig7} (a) shows the energy injection rate, $\dot E$, as a function of the sonic Mach number, $M_S$, for all the runs. Figure \ref{fig:fig7} (b) shows the mean total compressible energy cascade rate, $\langle \varepsilon_C \rangle$, computed using the exact third-order law, plotted as a function of the energy injection rate, $\dot{E}$. In most cases, the cascade rate is found to be slightly lower than the injection rate. This discrepancy is likely due to the fact that $\varepsilon_C(\ell)$ is defined as a scale-dependent correlation function, and the estimate of $\langle \varepsilon_C \rangle$ relies on averaging over a finite portion of the inertial range. Such averaging may partially underestimate the true mean value. Despite this, a strong correlation is observed between the cascade and the injection rate.}

\section{DISCUSSION AND CONCLUSIONS}
\label{sec:conclusions}
We have performed three-dimensional direct numerical simulations to study the compressible and isothermal energy cascade rates in fully developed MHD turbulence across a broad range of plasma compressibility and magnetic guide field strengths. Specifically, we used the compressible and isothermal \citep{A2017a,B2013} exact relations to compute both the flux and non-flux contributions to the energy cascade rates. Our numerical results reveal a clear dependence of the energy cascade on plasma compressibility and the strength of the mean magnetic field. Moreover, we observed a consistent dominance of flux contributions over non-flux contributions when the signed cascade rate is computed in these compressibility levels.


An estimation of the compressible energy cascade rate was obtained by \citet{A2020} using PSP, THEMIS and MAVEN observations in the pristine solar wind. The authors computed both incompressible and compressible cascade rates for hundreds of events at different heliocentric distances, ranging from  0.2 au to 1.7, finding a moderate increases in $\varepsilon_C$ with respect to $\varepsilon_I$, depending on the compressibility and distance to the Sun. Our numerical results show that both compressible and incompressible flux terms increase as we increase the compressibility, with this effect being more pronounced as we increase the magnetic guide field strength. Furthermore, we observe that the purely compressible flux term, Eq.~\eqref{f2c}, is approximately an order of magnitude smaller than the Yaglom-like term, \eqref{f1c}, in agreement with previous observation and numerical results \citep{B2016c,H2017a,A2018,A2020,B2023}. In particular, \citet{B2023} analyzed the signed and absolute values of incompressible and compressible MHD energy cascade rates in the solar wind at different heliocentric distances using PSP data and exact relations for compressible isothermal \citep{A2017a} and polytropic \citep{S2022} MHD turbulence. Their observational results are fully consistent with our numerical results: a clear increase of the absolute value of $\langle |\varepsilon_C| \rangle$ and $\langle |\varepsilon_I|\rangle$ as the heliocentric distance decreases (i.e. as the magnetic guide field increases) and as the density fluctuations increase in the plasma. 

\citet{A2018b} numerically investigated compressible and isothermal MHD turbulence in subsonic regimes (up to $M_S=0.5$), finding that compressible flux term are dominant in the inertial range. Their results also revealed a weak dependence of the flux terms on the magnetic guide field, in contrast to the non-flux terms. Our findings not only align with but also extend their conclusions to supersonic regimes. Specifically, in our runs within the subsonic ranges ($M_S=0.5$ with varying $M_A$), the signed increments method shows a flux dominance by over an order of magnitude, as expected. In comparison, results from the absolute increments method show comparable contributions even in this subsonic and quasi incompressible regime. These findings suggest that the signed increments method provides a more reliable and physically consistent approach for analyzing compressible energy cascade rates. A interesting result we are obtaining is $\expval{|F_{2C}|}$ reaches its maximum relative to the incompressible flux $\expval{|F_{1C}|}$ at intermediate levels of compressibility, regardless of the mean magnetic field strength (see Fig.~\ref{fig:fig5}). In isothermal hydrodynamic turbulence, at higher Mach numbers, the Yaglom-like flux component $\expval{|F_{1C}|}$ is expected to dominate in the inertial range, as was reported by \citet{K2013}. While this dominance is typically associated with the limit $c_S\to 0$, in our simulations with a constant $c_S=1$, it instead arises from the increase in $u^{\text{rms}}$. Conversely, at low Mach numbers, the expected behavior of $\expval{|F_{2C}|}$ approaches the incompressible limit, where $\delta \rho \to 0$, as expected. Moreover, \citet{F2022} investigated the compressible exact relation in Hall MHD turbulence in the subsonic regime, concluding that purely compressible flux components are negligible compared to the Yaglom-like flux term, consistent with the trend observed in our results. 

The observed peak in the energy cascade rates at intermediate polar angles is consistent with prior findings in anisotropic turbulence studies such as \citet{Jiang2023}. That study suggests that energy transfer rates vary significantly with polar angle due to differences in the effective inertial range across scales, as discussed by \citet{Wang2022}. Our results support this notion, indicating that the cascade is most efficient at intermediate angles. One possible explanation is that, at these angles, energy transfer benefits from both strong perpendicular cascade dynamics (which dominate in quasi‐2D turbulence) and contributions from parallel transfer mechanisms. This optimal alignment may maximize the rate at which energy is redistributed across scales before dissipation becomes dominant.

As we previously discussed, to calculate the energy cascade rates within the MHD scales using the exact law \eqref{exact_law}, we have employed two approaches: one based on signed increments of the variables and another using absolute increments. While the absolute value approach improves statistical convergence \citep[e.g.,][]{P2009}, it tends to overestimate the non-flux contributions relative to the flux terms \citep[see also,][]{B2023, A2019a, V2024}. By contrast, the signed increment method, applied along different directions in the simulation domain, consistently highlights the dominance of the flux contributions across all levels of compressibility and mean magnetic field strengths explored in this work. However, it is important to note that this trend may not hold in localized regions of the plasma, such as current sheets or filamentary structures, where the hypothesis for the exact law derivation could not hold and presumably non-flux contributions could become dominant \citep{M2022,F2020}. Specifically, \citet{F2020} have shown that in supersonic turbulence, a dominant compressible (non-flux) component can emerge, marking a fundamental departure from the traditional flux-driven cascade \citep{K1941a,P1998a,K2013}, which is typically associated with filamentary structure formation. {\color{blue} In this context, for sonic Mach numbers $M_S > 6$, we expect the non-flux terms to become increasingly significant, as suggested by the trend observed in Figure~\ref{fig:fig6}, where the ratio of non-flux to flux contributions grows with increasing $M_S$. Figure~\ref{fig:fig7} supports the cascade picture by showing correlation between the estimated total compressible energy cascade rate, $\langle \varepsilon_C \rangle$, and the energy injection rate. However, a detailed investigation of the compressible energy cascade rate within specific plasma structures, both at MHD and kinetic (ion) scales, remains an important direction for future work.}

%
%
%
%
%
%
%
%
%
%
%

\begin{acknowledgments}
G.J.A., P.D. and N.A.~thank Fouad Sahraoui for useful discussions. G.J.A., P.D. and N.A.~acknowledge financial support from the following grants: PIP Grant No.~11220200101752, UBACyT Grant No.~20020220300122BA and Redes de Alto Impacto REMATE from Argentina. N.A. acknowledges the International Space Science Institute (ISSI) in Bern. DC and BR for this work was provided in part by the National Science Foundation
under Grant AAG-1616026.  Simulations were performed on \emph{Stampede2}, part of the Extreme Science and Engineering Discovery Environment \citep[XSEDE;][]{towns2014}, which is supported by National Science Foundation grant number ACI-1548562, under XSEDE allocation TG-AST140008. 
\end{acknowledgments}


\bibliography{cites}

\end{document}